\documentclass[twocolumn,preprintnumbers,aps]{revtex4}

\usepackage{latexsym}
\begin{document}
\bibliographystyle{plain}

\preprint{FERMILAB--CONF--02/048--T}

\title[Why We Are Here]{Why We Are Here}



\newcommand{\m}{\hbox{ m}}

\author{Chris Quigg}
\email[E-mail: ]{quigg@fnal.gov}
\thanks{Fermilab is operated by Universities Research Association Inc.
under Contract No.  DE-AC02-76CH03000 with the United States Department
of Energy.}
\affiliation{Fermi National Accelerator Laboratory \\ P.O. Box 500, Batavia IL 60510 USA}


\date{\today}

\begin{abstract}
Opening talk at \textit{Snowmass 2001:} a summer study on the future of particle physics.
\end{abstract}

\maketitle

\section{Welcome!}
On behalf of the American Physical Society, the Division of
Particles and Fields, and our partners in the Division of Physics of Beams,
it is my great pleasure to welcome you to \textit{Snowmass 2001:} a summer study on the
future of particle physics.  We are off to a wonderful start, with more than 1000
participants already registered. Thanks in part to generous support from the National
Science Foundation, the DPB, and the DPF, more than eighty students have come to
Snowmass. The students form part of a contingent of more than 200 of our colleagues
who are, by their own admission---no questions asked!---``young.'' 

I'd also like to add a special welcome, and hearty thanks, to the more
than 150 participants who have come to Snowmass from outside the United
States. We look forward to drawing on your expertise and your
perspectives as we try to shape the future of our subject. It is plain
to everyone, I hope, that international cooperation will make many more
futures possible: the choices we make can be when and where, not just
yes to this and no to that.

The exceptional infrastructure and stimulating ambience we find here at
Snowmass would not have been possible without the enthusiastic
support---material and otherwise---of our sponsors. Our funding
agencies, the United States Department of Energy, the National Science
Foundation, and NASA, have provided resources and encouragement. The
APS Division of Particles \& Fields and Division of Physics of Beams
made substantial contributions to the outreach effort---which also
benefited from very considerable donations from private foundations
that wish to remain anonymous---and to student support. Another of our
professional organizations, the Nuclear and Plasma Sciencies Society of
the IEEE, has organized and financed an all-star ``technology
emphasis'' that will run throughout the three weeks of Snowmass 2001.
We're also extremely grateful for financial support and the show of
solidarity from ten laboratories engaged in particle physics research
in the United States: Argonne National Lab, Berkeley Lab, Brookhaven
National Lab / Brookhaven Science Associates, Cornell University / LNS
/ Wilson Synchrotron Lab, Fermilab / Universities Research Association,
Jefferson Laboratory / SURA, Lawrence Livermore National Laboratory,
Los Alamos National Laboratory, Oak Ridge National Lab / Spallation
Neutron Source, Stanford Linear Accelerator Center / Stanford
University. We thank all these institutions for their commitment to
particle physics, and to Snowmass 2001, and we salute the organizations
behind our great laboratories for their stewardship.

The stimulating and inclusive program of activities we will enjoy
during the next three weeks is the creation of the Snowmass Organizing
Committee. It has been a joy to work with these thoughtful, creative
people: they have given enormous amounts of time, energy, and goodwill
to the task of building a framework in which we can explore the whole
range of scientific opportunities and confront many kinds of issues
that will influence the conduct of particle physics in the future.
Because he cannot be here to see the fruits of his labors, I especially
want to call attention to the energetic leadership and great wisdom of
Ron Davidson, the DPB co-chair of the organizing committee. Ron brought
wonderful ideas and a confident serenity to the committee's work. Ron
and I join in thanking our colleagues Sally Dawson (BNL), Paul Grannis
(Stony Brook), David Gross (ITP/UCSB), Joe Lykken (Fermilab), Hitoshi
Murayama (Berkeley), Ren\'e Ong (UCLA), Natalie Roe (LBNL), Heidi
Schellman (Northwestern), and Maria Spiropulu (Chicago) from the DPF
side; and Alex Chao (SLAC), Alex Dragt (Maryland), Gerry Dugan
(Cornell), Norbert Holtkamp (SNS), Chan Joshi (UCLA), Thomas Roser
(BNL), Ron Ruth (SLAC), John Seeman (SLAC), and Jim Strait (Fermilab) from
the DPB side.

As we begin our adventure, I want also to thank the local organizing
committee, whose work you will see all around you over the next three
weeks, and our many working-group convenors from around the world, who
have responded so enthusiastically to their charges and will give 
definition to our work together.

\section{The  State of Particle Physics}
The physics curriculum in the 1898--99 University of 
Chicago catalogue begins with a very triumphalist Victorian preface~\cite{sbt}:
\begin{quote}
    ``While it is never safe to affirm that the future of the Physical
    Sciences has no marvels in store even more astonishing than those of
    the past, it seems probable that most of the grand underlying
    principles have been firmly established and that further advances are
    to be sought chiefly in the rigorous application of these principles
    to all the phenomena which come under our notice \ldots .  An eminent
    physicist has remarked that the future truths of Physical Science are
    to be looked for in the sixth place of decimals.''
\end{quote}
As the ink was drying on these earnest words, R\"{o}ntgen discovered x
rays and published the epoch-making radiograph of his wife's hand,
Becquerel and the Curies explored radioactivity, Thomson discovered
the electron and showed that the ``uncuttable'' atom had parts, and
Planck noted that anomalies in the \textit{first} place of the
decimals required a wholesale revision of the physicist's conception of
the laws of Nature.

We have the benefit of a century of additional experience and insight,
but we are not nearly so confident as our illustrious Victorian ancestors were
 that we have uncovered ``most of
the grand underlying principles.''  Indeed, while we celebrate the
insights codified in the \textit{standard model of particle physics}
and look forward to resolving its puzzles, we are increasingly
conscious of how little of the physical universe we have experienced and explored. 
Future truths are still to be found in precision measurements, but the
century we are leaving has repeatedly shown that Nature's marvels are
not limited by our imagination.  Exploration can yield surprises that
completely change what we think about---and how we think.
\subsection{A Decade of Discovery Ahead}
Over the next decade, we  look forward to an avalanche of 
experimental results that have the potential to change our view of the
fundamental particles and their interactions in very dramatic ways.  A
special preoccupation for me is the search and study of the Higgs
boson; this is really shorthand for a thorough exploration of the
1-TeV scale, which will elucidate the mechanism of electroweak
symmetry breaking.  We can also expect wonderful progress in flavor
physics: the detailed study of \textsf{CP} violation in the $B$
system, dramatically increased sensitivity in the exploration of rare
decays of $K$ and $D$ mesons, and pinning down the nature of neutrino
oscillations.  Maybe we will at last see a \textsf{CP}-violating
permanent electric dipole moment of the neutron.  Run II of the
Tevatron will give us our first opportunity to regard the top quark as a
tool, and not only as an object of desire.  Although the
interpretation of heavy-ion collisions at RHIC and the LHC promises to
be challenging, the heavy-ion colliders offer a real chance to
discover new phases of matter and enrich our understanding of QCD. 

On
many fronts, we are taking dramatic steps in energy and sensitivity
that will help us \textit{explore:} extra dimensions, new dynamics,
supersymmetry, and new kinds of forces and constituents might show
themselves.  (I'm conflicted about whether I'd like to see them all at
once, or in easy-to-understand installments!)

Experiments that use natural sources also hold great promise for the
decade ahead.  We suspect that the detection of proton decay is only a
few orders of magnitude away in sensitivity.  Astronomical observations
should help to tell us what kinds of matter and energy make up the
universe.  The areas already under development---if not
exploitation---include gravity wave detectors, neutrino telescopes,
cosmic microwave background measurements, cosmic-ray observatories,
$\gamma$-ray astronomy, and large-scale optical surveys. Indeed, the
whole complex of experiments and observations we call
astro/cosmo/particle physics should enjoy a golden age.

Here at Snowmass, we will have the opportunity to consider
 many imaginative ideas for instruments and
experiments that lie beyond our current horizon. Although theoretical
speculation and synthesis is valuable and necessary, we cannot advance
without new observations.  The experimental clues needed to answer
today's central questions can come from experiments at high-energy
accelerators, experiments at low-energy accelerators and nuclear
reactors, experiments with found beams, and deductions from
astrophysical measurements.  Past experience, our intuition, and the
current state of  theory all point to an indispensable role for
accelerator experiments.

The opportunities for accelerator science and technology are
multifaceted and challenging, and offer rich rewards for particle
physics.  

One line of attack consists in refining known technologies to
accelerate and collide the traditional projectiles---electrons,
protons, and their antiparticles---pushing the frontiers of energy,
sensitivity, and precise control.  The new instruments might include
brighter proton sources; very-high-luminosity $e^{+}e^{-}$
``factories'' for $B$, $\tau$ / charm, $\phi$, \ldots; a Tevatron
``Tripler'' based on high-field magnets; cost-effective hadron
colliders beyond the LHC at CERN, represented by the Super-LHC and Very
Large Hadron Collider initiatives; and $e^{+}e^{-}$ linear colliders.

A second approach entails the development of exotic acceleration
technologies for standard particles: electrons, protons, and their 
antiparticles.  We don't yet know what
instruments might result from research into new acceleration methods,
but it is easy to imagine dramatic new possibilities for particle
physics, condensed matter physics, applied science, medical
diagnostics and therapies, and manufacturing, as well as a multitude of
security applications. A teach-in on July 5 will explore opportunities to become
involved in research on advanced acceleration methods.

A third path involves the exploration of exotic particles for
accelerators and colliders to expand the experimenter's 
armamentarium.  Muon storage rings for neutrino factories,
$\mu^{+}\mu^{-}$ colliders and $\gamma\gamma$ colliders are all under 
active investigation, and each of these would bring remarkable new 
possibilities for experiment.

Finally, let us note the continuing importance of enabling 
technologies: developing or domesticating new materials, 
new construction methods, new instrumentation, and new active 
controls.  I call your attention to the IEEE Nuclear and Plasma Sciences 
Society's program of Technology Short Courses and Lunchtime Lectures, beginning
on July 5. 

To a very great extent, the progress of particle physics has been
paced by progress in accelerator science and technology.  A renewed
commitment to accelerator research and development will
ensure a vigorous intellectual life for accelerator science and lead
to important new tools for particle physics and beyond.

\subsection{Creating the Future}
 
Now, the decade of discovery won't happen automatically.  Many of our 
goals are difficult, and timely success is in doubt for many 
experiments.  We must push hard to prepare the instruments, and get to 
the answers.

The glorious future of new machines and new experiments that lies beyond 
the established program also won't happen by itself.  We have, I 
think, come to the collective realization that we must do more to 
prepare alternative futures by creating a rich and organic program of 
accelerator research.  

We're also challenged by our success: \textit{the scope of our 
science has grown, but funding has not.}  Within our own extended 
family and beyond, we must do more to convey the urgency and 
importance of the new scientific opportunities, and fashion a program 
that we can carry out that includes the right measure of scale 
diversity to ensure a healthy intellectual ecosystem.

Many individuals in our community, all the major laboratories, and many
university departments work energetically---and effectively---in
outreach and educational activities, conveying the excitement and
substance of science---and particle physics---to students, and to the
general public. But we are not doing everything we might to promote
scientific literacy, to inspire the next generation of scientists,
engineers, and technologists, and to report to our patrons---our fellow
citizens in government and in the general public---our hopes and
dreams, our triumphs and challenges. We can communicate much more
effectively the wonders of our science. It is in our interest to do so,
and it is our obligation to those who support our work. 

To show how seriously we take the need to present our science to
others, we have expanded the usual technical program of a Snowmass summer study
to include a vigorous and diverse program of outreach and educational
activities---right here in the Roaring Fork Valley. The presence of
Quarknet teachers, the Science Weekend extravaganza on the Snowmass
Mall next weekend, and a schedule of public lectures in Aspen,
Snowmass, and Carbondale are only part of the story. I hope that many
of you will participate in the outreach program, and that all of you
will take time to see what your colleagues are doing and to think about
what more you might be doing at home. 

I also call your attention to the Communications workshops scheduled
this week and next, and to the ``Working with Governments'' forum near
the end of the summer study. To add to the literature that presents the
achievements and aspirations of particle physics to a broad audience,
the Division of Particles and Fields is preparing an illustrated
thematic survey entitled \textit{Quarks Unbound}.

\subsection{What We Need to Know}
Plans that proceed from broad scientific goals to specific questions
and then to instruments and technology development have been used to
excellent effect by the National Cancer Institute and by NASA. In
organizing my thoughts about our future, I find it useful to consider
the agenda of particle physics today under a few broad rubrics.  

\vspace*{3pt}\noindent\textit{Elementarity.} Are the quarks and leptons structureless, or
will we find that they are composite particles with internal
structures that help us understand the properties of the individual
quarks and leptons?

\vspace*{3pt}\noindent\textit{Symmetry.} One of the most powerful lessons of the modern
synthesis of particle physics is that (local) symmetries prescribe
interactions.  Our investigation of symmetry must address the question
of which gauge symmetries exist (and, eventually, why).  We have
learned to seek symmetry in the laws of Nature, not necessarily in
the consequences of those laws.  Accordingly, we must understand how
the symmetries are hidden from us in the world we inhabit.  For the
moment, the most urgent problem in particle physics is to complete our
understanding of electroweak symmetry breaking by exploring the 1-TeV
scale.  This is the business of the experiments at LEP2, the Tevatron
Collider, and the Large Hadron Collider.

\vspace*{3pt}\noindent\textit{Unity.} In the sense of developing explanations that apply not
to one individual phenomenon in isolation, but to many phenomena in
common, unity is central to all of physics, and indeed to all of
science.  At this moment in particle physics, our
quest for unity takes several forms.

First, we have the fascinating possibility of gauge coupling
unification, the idea that all the interactions we encounter have a
common origin and thus a common strength at suitably high energy.

Second, there is the imperative of anomaly freedom in the electroweak
theory, which urges us to treat quarks and leptons together, not as
completely independent species.  Both of these ideas are embodied, of
course, in unified theories of the strong, weak, and electromagnetic
interactions, which imply the existence of still other forces---to
complete the grander gauge group of the unified theory---including
interactions that change quarks into leptons.

The third aspect of unity is the idea that the traditional distinction
between force particles and constituents might give way to a unified
understanding of all the particles.  The gluons of QCD carry color
charge, so we can imagine quarkless hadronic matter in the form of
glueballs.  Beyond that breaking down of the wall between messengers
and constituents, supersymmetry relates fermions and bosons.

Finally, we desire a reconciliation between the pervasive outsider,
gravity, and the forces that prevail in the quantum world of our
everyday laboratory experience.

\vspace*{3pt}\noindent\textit{Identity.} We do not understand the physics that sets quark
masses and mixings.  Although we are testing the idea that the phase
in the quark-mixing matrix lies behind the observed \textsf{CP}
violation, we do not know what determines that phase.  The
accumulating evidence for neutrino oscillations presents us with a new
embodiment of these puzzles in the lepton sector.  At bottom, the
question of identity is very simple to state: What makes an electron
and electron, and a top quark a top quark?

\vspace*{3pt}\noindent\textit{Topography.} ``What is the
dimensionality of spacetime?''  tests our preconceptions and unspoken
assumptions.  It is given immediacy by recent theoretical work.  For
its internal consistency, string theory requires an additional six or
seven space dimensions, beyond the $3+1$ dimensions of everyday
experience.  Until recently it has been presumed that the extra
dimensions must be compactified on the Planck scale, with a
stupendously small compactification radius $R \simeq
M_{\mathrm{Planck}}^{-1} = 1.6 \times 10^{-35}\m$.
Part of the vision of string theory is that what goes on in even such
tiny curled-up dimensions does affect the everyday world: excitations
of the Calabi--Yau manifolds determine the fermion
spectrum.

We have recognized recently that Planck-scale compactification is 
not---according to what we can establish---obligatory, and that current
experiment and observation admit the possibility of dimensions not 
navigated by the strong, weak, and electromagnetic interactions that 
are almost palpably large.  A whole range of new experiments will help 
us explore the fabric of space and time, in ways we didn't expect just 
a few years ago.
\section{Some Goals for Snowmass 2001}
All of you have arrived in Snowmass with ambitious plans for the next
three weeks.  The charges to the twenty-seven working groups give a detailed picture of
the Organizing Committee's vision of what we can achieve, but here are the
high-level goals I hope we can attain together:

\noindent$\rhd$ Survey our aspirations for particle physics over 30
years.  

\noindent$\rhd$ Assess the current state of development of accelerator protoprojects
and advanced accelerator research, and understand the investment we
must make (financial and human capital) to bring the most promising
lines to maturity.  Here at Snowmass, the Division of Physics of Beams will
be completing a very broad look at our opportunities and needs for Accelerator R\&D. 

\noindent$\rhd$ Look beyond our immediate goals for measurements and searches to
contemplate the shape of a more complete, more ambitious theoretical
framework.  How should theoretical vision shape our experimental goals?

\noindent$\rhd$ Examine the importance of scale diversity for a healthy and productive
future.

\noindent$\rhd$ Educate ourselves about the full range of possibilities before us.  We
must know enough to judge critically, to improve the arguments, to
articulate our goals effectively.  We're very grateful to the senior
scientists who have agreed to serve as HMOs---high-minded outsiders---in experimental
working groups E1 -- E6. Their assignment is to act as friendly skeptics,
probing and strengthening arguments, and---by the example of their time and
effort---to lead many others to view Snowmass 2001 as a forum for engaging with
the ideas and aspirations of others.

\noindent$\rhd$ Listen carefully to our young colleagues, who will help create our
common futures.   We'll have a formal opportunity to hear from the next generation
at the Young Physicists Forum on July 17, so mark that date on your calendar.

\noindent$\rhd$ Take advantage of opportunities to interact with the HEPAP Subpanel.
Technical work carried out at Snowmass will undergird the
recommendations the subpanel makes.

\noindent$\rhd$ Consider the international dimensions of what we hope to achieve.
It will be a particular pleasure to
welcome the international lab directors, and I call your attention to an 
evening discussion this week on the issues of a global accelerator network.
During the final week of Snowmass 2001, we'll have reports from the European
 and Japanese high-energy physics planning Committees

I believe we must articulate a comprehensive vision of particle physics
(and the sciences it touches) to make our case effectively to
ourselves, to other scientists, and to society at large.   At the same
time, we have a special responsibility to examine the prospects for the
most ambitious accelerators, which are major drivers of our scientific
progress.   If we judge the science to be rich, and if we can make the
cost and technical risk attractive, we will want to pursue all the
leading possibilities: linear colliders, hadron colliders reaching far
beyond the TeV scale, muon storage ring, and muon collider.   The
vision we present should include the scientific promise of all these
instruments, and a strategy for deciding what, where, and when that
includes the organic R\&D investment we will need to evolve the right
set of instruments to serve our science.

\noindent$\rhd$ Thanks to the work of many people, the moment is upon us to probe,
shape, and judge the idea of a linear collider as a possible next big
step for particle physics. \textit{Evaluating a linear collider and
working to define a scientifically rich, technically sound, fiscally
responsible plan is a homework problem for the entire community.} 
Everyone must come to an informed judgment.   

Beyond the technical issues, please think about how to make our dreams
happen.  Creating a future is not accomplished when we draw our
individual conclusions or read the subpanel recommendations.

Welcome to Snowmass 2001!
Your passion, energy, creativity, and commitment will change the world.


\end{document}